# Thermally-reconfigurable metalens

*Anna* Archetti[1,2,†], *Ren-Jie* Lin[1,†], *Nathanaël* Restori[1], *Fatemeh* Kiani[1], *Ted V.* Tsoulos[1] and *Giulia* Tagliabue[1,*]

[1]Laboratory of Nanoscience for Energy Technologies (LNET), STI, École Polytechnique Fédérale de Lausanne, 1015 Lausanne, Switzerland

[2]Enlightening Brain Mechanisms, Neuroscience, Department of Biomedical Sciences, University of Padua, 35131 Padova, Italy

† equal contribution

* corresponding author: giulia.tagliabue@epfl.ch

**Abstract**

Thanks to the compact design and multi-functional light-manipulation capabilities, reconfigurable metalenses, which consist of arrays of sub-wavelength meta-atoms, offer unique opportunities for advanced optical systems, from microscopy to augmented reality platforms. Although poorly explored in the context of reconfigurable metalens, thermo-optical effects in resonant silicon nanoresonators have recently emerged as a viable strategy to realize tunable meta-atoms. In this work, we report the proof-of-concept design of an ultrathin (300 nm thick) and thermo-optically reconfigurable silicon metalens operating at a fixed, visible wavelength (632 nm). Importantly, we demonstrate continuous, linear modulation of the focal-length up to 21% (from 165 µm at 20°C to 135 µm at 260°C). Operating under right-circularly polarized light, our metalens exhibits an average conversion efficiency of 26%, close to mechanically modulated devices, and has a diffraction-limited performance. Overall, we envision that, combined with machine-learning algorithms for further optimization of the meta-atoms, thermally-reconfigurable metalenses with improved performance will be possible. Also, the generality of this approach could offer inspiration for the realization of active metasurfaces with other emerging material within field of thermo-nanophotonics.

**Introduction**

Optical metasurfaces are two-dimensional (2D) arrangements of meta-atoms with sub-wavelength spacing that are engineered to precisely control and manipulate the properties of a light beam, such as its phase, amplitude, and polarization. Their evolution in the last two decades has enabled the miniaturization of numerous optical components, including holographic optical elements[1–3], beam deflectors[2,4], and flat lenses[5–8]. More recently, reconfigurable metasurfaces have enabled active control of the optical properties revolutionizing the design, functionality and application domains of optical components and devices[9–12]. Phase-change materials[9,13–15], stretchable substrates[16–18], strain-field engineering[19], thermo-optical effects[20–23], and free-carrier modulation[10,24–26] are some of the engineering strategies used to achieve reversible changes of the metasurface function.

Reconfigurable (varifocal) metalenses (R-MLs)[15,16,21,23,25], in particular, have attracted increasing attention thanks to their enabling potential for tunable optics in microscopy systems[27], depth sensor devices[28,29] as well as virtual and augmented reality platforms[30]. Moreover, these R-MLs can uniquely combine focal length modulation with complex functionalities, such as multi-wavelength[31] operation, spectroscopy and polarization routing[32], that are critical for compact and smart imaging devices/sensors.

The realization of *continuously* reconfigurable MLs, however, presents significant challenges. Indeed, in order to generate a converging wavefront with a desired focal length, the nano-scatters constituting a ML must introduce a prescribed (parabolic) phase delay at each position along the ML radius. (**Figure 1a-c**). To dynamically change the focal length, a distinct phase shift must be achieved at each lattice site. Thus, R-MLs require the non-trivial realization of both a parabolic spatial phase-profile and a spatially-varying phase shift (**Figure 1d**).

To date, stretchable substrates[33] and strain-field engineering[34], which entail a mechanical modification of the ML structure, have demonstrated excellent reconfiguration capabilities with good efficiency (up to 30% and 60% respectively) and large focal length tuning (up to 20% and 100% respectively). Yet, there is a growing interest for non-mechanical modulation approaches such as phase-change materials, thermo-optical effects[35] as well as tuning of free-carriers and exciton-resonances[36]. Indeed, these designs can offer unique opportunities in terms of fast modulation speed, including ultra-fast all-optical control[37], and reduced device thickness, i.e. ultra-thin lenses[38]. Furthermore, achieving the desired modulation using a single uniform external control of the local optical properties (e.g. temperature or electrical potential), would greatly simplify device design, fabrication and integration, accelerating the deployment in real-word components.

Thermo-optic effects represent an attractive approach for the realization of dielectric R-MLs[23]. Indeed, they entail a continuous and smooth change in the material optical properties and are robust against thermal-cycling. Yet, the low-magnitude of typical thermo-optical coefficients is generally assumed to limit the applicability of this strategy. Interestingly, silicon nano-resonators [5,30] have been recently shown to exhibit pronounced shifts in their optical resonances (**Figure 1b**) both under external heating[20,22] and for all-optical[39] modulation. Thus, by leveraging the amplification of thermo-optical effects by optical-resonance modes, silicon-based thermally-reconfigurable dielectric metalenses (TR-MLs) could become a competitive and CMOS-compatible solution. However, a viable design for such an ultra-thin, tunable ML with fixed operation wavelength in the visible regime has not been demonstrated yet.

Here we report a proof-of-concept design of an ultrathin (300 nm thick) and thermo-optically reconfigurable silicon ML operating at a fixed wavelength in the visible regime (632 nm). We demonstrate that, using thermo-optic effects, it is indeed possible to achieve continuous modulation of the focal-length beyond the depth-of-focus of the lens. Specifically, operating under right-circularly polarized light, our TR-ML exhibits a change of 21% in the focal length, with a linear shift from $165\ \mu m$ at $20°C$ to $135\ \mu m$ at $260°C$. The average conversion efficiency of the lens is 26%, close to the performance of mechanically modulated devices, while its Strehl ratio is 0.99, confirming a diffraction-limited performance. Importantly, in our design, we rely on a spatially-uniform temperature increase of the structure, overcoming the need for a spatially-varying modulation input and potentially enabling an all-optical photo-thermal modulation approach. Concurrently, in this work, we report an automatized methodology to design a reconfigurable metalens, compute its layout and verify the expected performance. Overall, although further optimization of the meta-atom design is needed to boost the performance of these components, our results demonstrate that TR-MLs can be a viable solution for active tuning of optical systems.

**Results**

*Thermally-reconfigurable metalens phase profile and choice of design parameters*

A TR-ML must be composed of an array of nano-scatters (meta-atoms - **Figure 1a, c**) that, at all temperatures $T$, satisfies a quadratic phase profile[40] along its radius $r$:

$$\phi(r, f(T)) = -\frac{2\pi}{\lambda}\left[\sqrt{r^2 + f^2(T)} - f(T)\right] \quad (1)$$

where $f(T)$ is the temperature-dependent focal length and $\lambda$ the wavelength of the incident light. Fixing the desired focal length at the initial temperature $T_0$, $f_0 = f(T_0)$, and requiring $f$ to be linearly dependent on the temperature $T$, it is possible to express $f(T)$ as:

$$f(T) = (T - T_0)\frac{\Delta f}{\Delta T} + f_0 \quad \begin{matrix} \Delta T = T_{\max} - T_0 \\ \Delta f = f(T_{\max}) - f_0 \end{matrix} \quad (2)$$

where $\Delta f$ is the desired focal-length variation achievable at the maximum temperature variation $\Delta T$. Using this assumption for $f(T)$, and omitting the dependence on the fixed parameters $T_0, \Delta T, f_0, \Delta f$, the phase can be rewritten as follows:

$$\phi(r, T) = \phi_0(r) + \Delta\phi(r, T) \quad (3)$$

where $\phi_0(r) = \phi(r, f_0)$ and the required phase shift at each temperature is:

$$\Delta\phi(r, T) = \phi(r, T) - \frac{2\pi}{\lambda}\left[\sqrt{r^2 + f_0^2} - f_0\right] \quad (4)$$

We observe that in Equation ( 3 ) the required temperature-dependent phase profile $\phi(r, T)$ has been decomposed into a temperature-independent initial phase profile, $\phi_0(r)$, and a *spatially-varying* phase shift, $\Delta\phi$, which depends on the temperature. Tuning of the phase shift at each lattice site can be achieved either by a structured external control[41] $T(r)$ that generates different modulation inputs along the radius, i.e. $\Delta\phi(T(r))$, or by using meta-atoms that respond differently to the same uniform external control[42] $T$, i.e. $\Delta\phi(r, T)$. In our system, the phase shift must be generated by thermo-optical effects in dielectric nanofins. Applying different temperatures to nano-structures that are closely spaced is very challenging. We thus pursue the second strategy, and we consider a spatially-uniform temperature input $T$ (**Figure 1a**). The analytical phase profile described in Equation ( 3 ) is uniquely defined once $\lambda$, $\Delta T$, $f_0$, $\Delta f$, the ML radius $R$ are defined. We note that for each set of these constraints, a specific phase shift range $\Delta^{\max}\phi$ must be attained with the meta-atoms (**Figure 1c**), where:

$$\Delta^{\max}\phi = \max\left(\Delta\phi(r, T_{max})\right) - \min(\Delta\phi(r, T_{max})) \quad (5)$$

Seeking a favorable compromise between tunability and transmission efficiency of the ML in the visible regime, we set $\lambda = 632\ nm$ as the operating wavelength (**Figure 1b**). We also considered a maximum temperature variation $\Delta T = 240°C$, compatible with our experimental capabilities ($T_{max} = 260°C$). Next, we quantified $\Delta f/f_0$ as a function of the lens numerical aperture (NA), its radius $R$ as well as $\Delta^{\max}\phi$ and we studied the maximum focal length variation versus the depth of focus (**Figure 1f**). From this detailed analysis we observed that, in percentage, $\Delta f/f_0$ decreases for larger $R$ and for smaller $\Delta^{max}\phi$. Aiming for a ML with a numerical aperture (NA) of at least 0.08, comparable with commercial ultra-compact objectives, and with a focal length variation at least equal to the depth of focus, we defined the target phase profile of the TR-ML using the following parameters: $f_0 = f(T_0) = 200\ \mu m$, $\Delta f = -45\ \mu m$ and $R = 15.75\ \mu m$. We thus obtain $NA(f_0) = 0.08$, $NA(f_0 + \Delta f) = 0.10$, $\Delta^{max}\phi = 102$ deg, the average depth of focus $z_0 \sim 40\ \mu m$ and the linear increase of the focal length with temperature is $\Delta f/\Delta T \sim 1.9\ \mu m/°C$. The ideal ML phase profile and required phase change obtained for the chosen set of parameters are shown in **Figure 1d-e**.

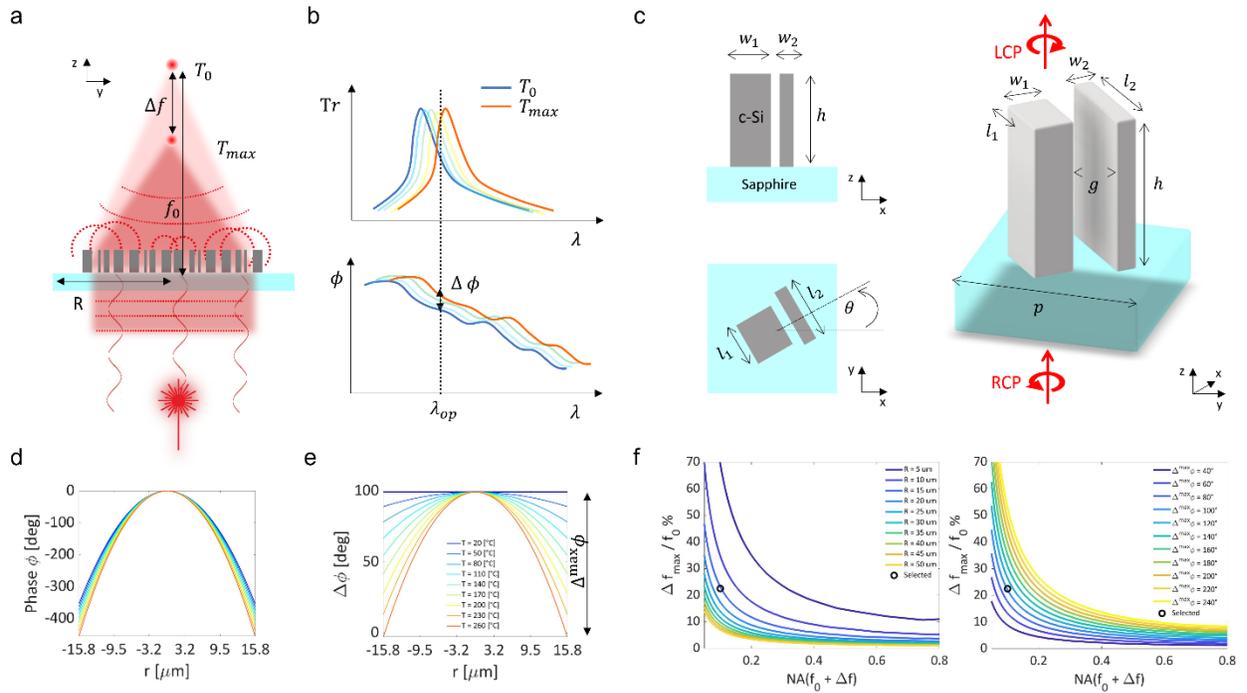

**Figure 1 – Nanoresonator-based thermally-reconfigurable phase control: principle and analytical phase design. a** Schematic illustration of a thermally reconfigurable metalens. Each nano-structure exhibits a resonance mode that locally induces a phase delay. All together these meta-atoms shape a converging wavefront with focal length $f_0$, which can be modulated with the temperature thanks to thermo-optical effects (schematic shows a decrease in focal length). **b** Schematic illustration of the thermo-optical effect. Thermally-induced changes of the refractive index cause a shift of the resonance mode of a meta-atom (top). The associated change in phase delay results in a phase shift $\Delta\phi$ at the operating wavelength $\lambda_{op}$ (bottom). **c** Lateral and top views of a nanofin (left) as well as its 3D rendering, indicating also the illumination conditions (right). ($l_1$, $w_1$, $l_2$, $w_2$), $gap$, $h$, $p$ and $\theta$ are the width and the length of the nanopillar 1 and 2, the gap between the two nano-pillars, the height of the nano-pillars corresponding to the initial c-Si film thickness, the nano-fin lattice period and the angle of rotation of the nanofin structure respectively. $g = 0.060\ \mu m$, $h = 0.300\ \mu m$, $p = 0.350\ \mu m$. **d** Required phase profile at different temperatures to satisfy: initial focal length $f_0 = 200\ \mu m$ at $T_0 = 20°C$; final focal length $f_0 + \Delta f = 155\ \mu m$ at $T_{max} = 260°C$. **e** Required phase shift $\Delta\phi(r,T)$ between the phase profile at $T_0 = 20°C$ and the phase at temperature $T$. **f** Numerical simulation of the maximum focal length variation over the initial focal length (%) as a function of the numerical aperture NA achievable with a maximum phase variation of $\Delta^{max}\phi \sim 100$ deg and with different ML radiuses (left). Maximum focal length variation over the initial focal length (%) as a function of NA achievable with a ML radius $R = 15.75\ \mu m$ and for different $\Delta^{max}\phi$ values (right). The chosen metalens design values are indicated with a black circle ($f_0 = 200\ \mu m$, $\Delta f = -45\ \mu m$ and $\Delta^{max}\phi = 102$ deg).

*Metalens design: silicon nanofin library and digitalization of the target phase profile*

The response of individual meta-atoms must be engineered such that their phase parameter space ($\phi$ and $\Delta\phi$) satisfies the requirements imposed by the target phase $\phi(r,T)$ and phase shift $\Delta\phi(r,T)$. Specifically, based on our design parameters discussed above, we require a phase range of $\phi = [0, 2\pi]$, a maximum phase shift range of $\Delta^{max}\phi \sim 100$ deg (**Figure 1d-e**) and a linear variation of the phase shift $\Delta\phi$ with temperature.

Non-resonant silicon nano-pillars, which only support waveguide-modes[5,43,44], are insufficient to realize an ultra-thin, sub-μm thick ML ($h = 300\ nm$) because, for $\Delta T = 240°C$, they can achieve a maximum phase shift $\Delta^{max}\phi \sim \frac{2\pi}{\lambda} h\,\Delta^{max}n \sim 12° \ll 100°$. Ultra-thin, resonant meta-atoms are instead expected to provide a wider phase space thanks to the amplification of thermo-optical effects by the different optical modes[39,45,46] (**Figure 1b**). Yet, simple geometries supporting a single resonance cannot produce a phase modulation range exceeding $\pi$ [47].

To introduce additional degrees of freedom for engineering the phase profile, we therefore adopted anisotropic meta-atoms (nanofin) composed of two silicon nano-pillars waveguides with different nano-pillar length and width that act as coupled waveguides (**Figure 1c**). Indeed, these anisotropic nanofins offer the opportunity to leverage the Pancharatnam-Berry phase[48,49] and obtain an additional $\pi$ phase accumulation[50] (see **Methods** for further details).

As shown in **Figure 1c** and **Figure 2**a, the engineered nanofins are characterized by a set of geometrical parameters $\# = (l_1,\ w_1, l_2,\ w_2)$, where 1 and 2 indicate the first and the second nano-pillar, respectively, and by their rotation angle $\theta$. The gap between the two nano-pillars ($g = 50\ nm$) and their height ($h = 300\ nm$) are fixed for all the nanofins (see **Methods** for further details). These are distributed with varying rotation angles in a sub-wavelength lattice on top of a sapphire substrate with a fixed period ($p = 350\ nm$).

We used numerical simulations (COMSOL Multiphysics®, see **Methods** for further details) to study the electromagnetic response of the nanofins. To obtain our nanofin library, we first performed a parameter sweep computing the transmission efficiency versus the maximum phase shift for every geometry # at a fixed angle $\theta = 0$ and for $\Delta T = 240°C$. Interestingly, there appears to be a trade-off between the maximum phase shift and the transmission efficiency. From this first study, we selected those geometries with an average transmission efficiency above 15% that could also cover the required $\Delta^{max}\phi \sim 100$ deg (**Figure 2a**). Next, for every selected geometry #, we performed a sweep over the rotation angle $\theta$ and input temperature $T$, obtaining the complete dataset of phase $\phi(\theta,\ T,\ \#)$, phase shift $\Delta\phi(\theta,\ \#)$ and transmission efficiency (**Figure 2b**). Finally, the nanofins must be distributed spatially based on the discretization of the analytical phase constraints. We thus constructed a routine that searched our dataset to identify, for every spatial position, the geometry (#) and rotation angle ($\theta$) that satisfied the required phase-delay at both $T = 20°C$ and $T = 260°C$ as well as the linearity condition. With this approach we obtained the phase profile digitalization and the resulting layout for an optimal design of our TR-ML (**Figure 2c-d**). In particular we observe that, leveraging the PB phase, only six nanofin geometries are sufficient to satisfy well all the constraints.

Overall, combining thermo-optical effects engineered silicon nanofins with a Pancharatnam-Berry (geometrical) phase provides a large range of phase and phase shift values enabling the design of a thermally reconfigurable metalens (TR-ML).

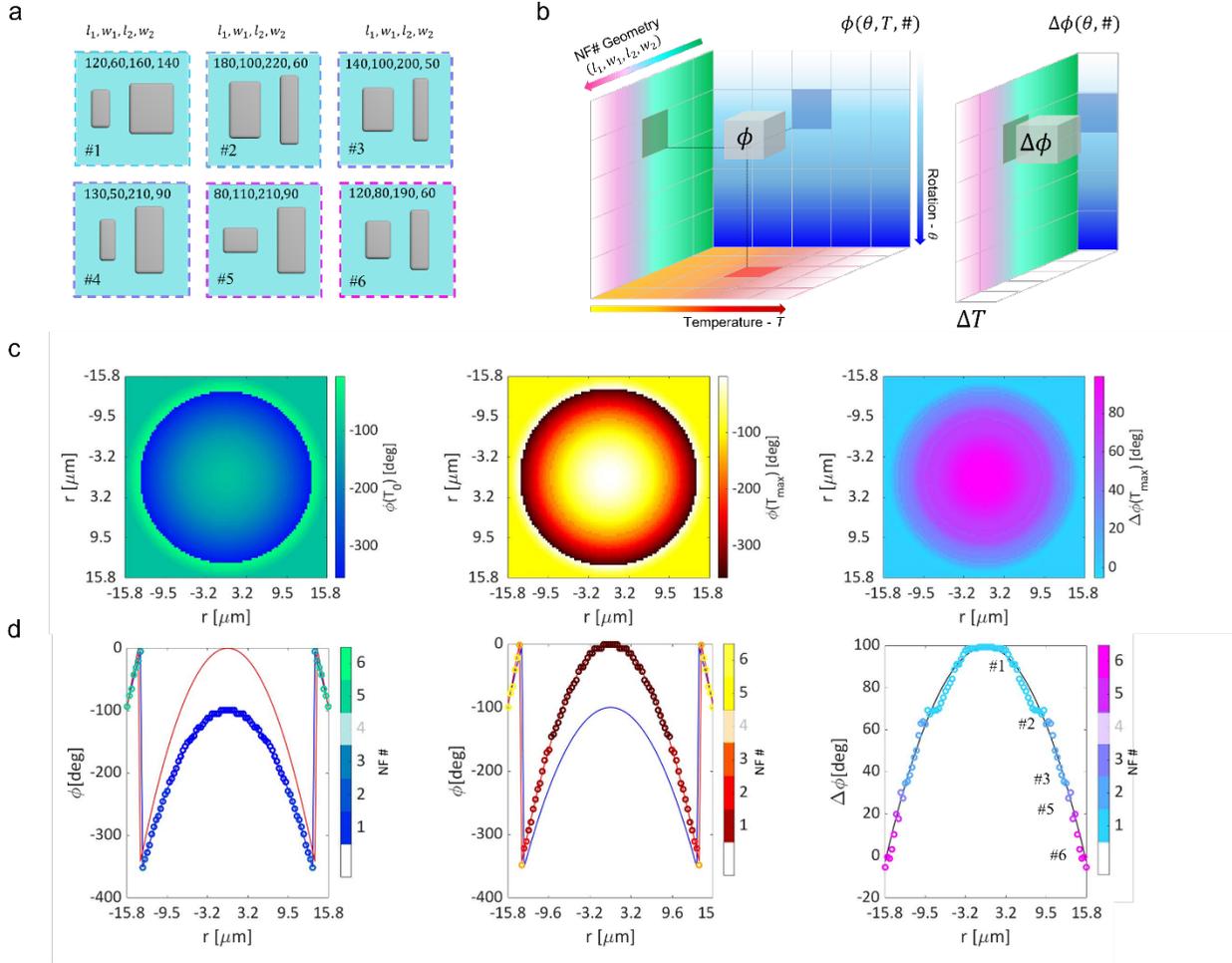

**Figure 2 – Study of the thermo-optic response of nanofins with different geometrical parameters and relative rotation. a** Schematic illustration of a unit nanofin NF # structure and its geometrical parameters; $(l_1, w_1, l_2, w_2)$, $g$, $h$, $p$ and $\theta$ are the width and the length of the nanopillar 1 and 2, the gap between the two nano-pillars, the height of the nano-pillars corresponding to the initial c-Si film thickness, the nano-fin lattice period and the angle of rotation of the nanofin structure respectively. $gap = 0.060\ \mu m$, $h = 0.300\ \mu m$, $p = 0.350\ \mu m$. **b** $\phi(\theta, T, \#)$ and $\Delta\phi(\theta, \#)$ are a 3D and 2D matrices representing the phase parameter space used for the wavefront design. **c left and center** 2D metalens phase profile at 20°C and 260°C respectively. **c right** Metalens phase shift between 260° and 20°C. **d left and center** 1D projection of the 2D plots **c left and center** respectively. Circle markers represent the actual ML phase value at a specific radial position overlapped to the theoretical phase values displayed with a continuous line (blue and red color indicates 20°C and 260°C respectively. **d right** 1D projection of the 2D plots in **c right.** Circle markers represent the actual ML phase shift value at a specific radial position overlapped to the theoretical phase shift values displayed with a continuous black line. Different gradual colors in plot **c and d** represent different nanofin (#1-6). Notice that the nanofin #4 has not been selected by the algorithm for the ML design. All the phase shift profiles are computed with respect to the initial temperature $T_0 = 20°C$. All the nanofin structures have been studied with COMSOL numeric simulations to retrieve their phase and transmission efficiency at different temperatures and rotation angles $\theta$.

*Thermally-reconfigurable metalens performance*

We simulated the propagation of a beam focused by our TR-ML, obtaining the three-dimensional intensity profile of the focused beam (see **Methods**). From a qualitative inspection of the intensity profiles along the propagation direction, we verified that the focal length decreases as the temperature increases (**Figure 3a-b**). Furthermore, we observed that both the point-spread-function (PSF) of the beam focus and the depth of focus become narrower at higher temperatures. Thus, the lens NA increases with temperature, as expected (**Figure 3c-d, f**). From a quantitative analysis we obtained that the ML focal length changes from $f(20°C) \sim 165 \, \mu m$ to $f(260°C) \sim 135 \, \mu m$ while the NA increases from $NA(20°C) = 0.10$ to $NA(260°C) = 0.12$ (**Figure 3e**).

Due to intrinsic diffraction of low Fresnel number lens[51], our ML, exhibits an unavoidable deviation of its focal length from the ideal focal length defined with geometric optics. In particular, the peak irradiance position $Z_p$ derived from diffraction theory lies at an average distance $\delta = Z_p - f \sim 27 \, \mu m$ from the target focal length $f$ over all temperatures. Nonetheless, the reported design satisfies all the tunability requirements. Indeed, the total focal length modulation, $\Delta f \sim 30 \, \mu m$, is comparable to the final depth of focus, $z_0 \sim 35 \, \mu m$. In addition, the modulation follows the desired linear trend. In fact, from a linear fit of the focal length versus temperature we obtain a slope $\Delta f / \Delta T \sim 1.3 \, \mu m/°C$, a root mean square error $RMSE = 3.7 \, um$ and a R-squared value $R^2 = 0.99$ (**Figure 3e**).

Importantly, from the deviation of the discretized phase profile from the analytical one, we obtain a ML Strehl ratio of $S \sim 0.99 > 0.8$, indicative of a diffraction-limited behavior of our design (see **Methods** for further details). Indeed, as shown in **Figure 3f**, the value of the full-width at half maximum of the Airy disk at the focal plane ($FWHM = 2\sqrt{2\ln(2)} * \sigma$) lies at less than $0.1 \, \mu m$ from the diffraction-limit one ($\sigma \sim 0.42 \, \lambda \, / \, 2NA$ in the Gaussian approximation). Finally, our TR-ML exhibits an average focusing efficiency equal to 26%, with an average deviation of 2% over all temperatures (**Figure 3g**). Although, this value is limited by the optical losses and reduced transmission (intrinsically limited up to 50% by the polarization conversion and by the conversion-efficiency of the nanofin meta-atoms), our result is comparable to that of mechanically-actuated reconfigurable metalenses[33,52,53].

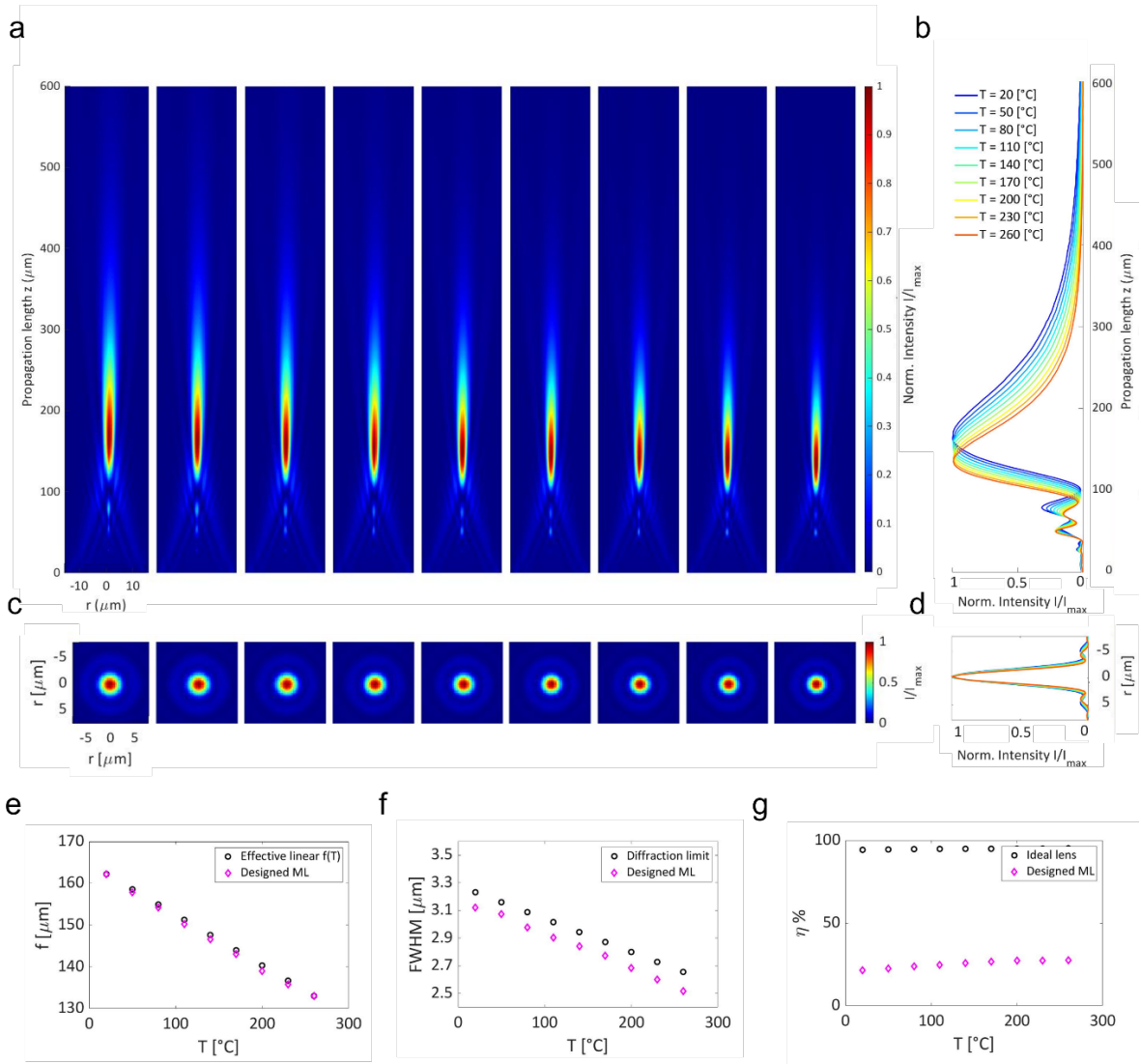

**Figure 3 – Thermally tunable metalens performance and deviation from an ideal lens. a** Beam propagation focused by the designed ML. 2D intensity profiles along the XZ propagation plane at increasing temperatures. **b** 1D projection of the intensity profiles along the Z propagation direction at increasing temperatures. **c** Point spread function (PSF) 2D profile at the focal plane (XY) at increasing temperatures. **d** 1D projection of the PSF at all simulated temperatures. All the intensity profiles are normalized by the maximum intensity value at each temperature. **e** Focal length $f$ at increasing temperatures. The focal length position of our designed metalens is compared with the focal length position extracted from the beam propagation of the ideal ML and with the effective linear focal length assumption. **f** PSF full width at half maximum (FWHM) of our designed ML at increasing temperatures. The FWHM of our designed ML is compared with the FWHM expected from an ideal ML and with the theoretical diffraction limit extracted from the Gaussian approximation of the Airy disk ($\sigma \sim 0.42\,\lambda\,/\,2NA$ and $FWHM = 2\sqrt{2\ln(2)} * \sigma$). **g.** Focusing efficiency $\eta$ of our designed metalens at increasing temperature. $\eta$ is computed as the transmitted field intensity over an integration area with radius $R_i = 17\,\mu m$ divided by the input field intensity at the ML plane ($z = 0\,\mu m$). The average transmission efficiency of our designed ML over all temperature is 26%. The efficiency of our designed ML is compared with the efficiency expected by an ideal ML and ideal lens. With "ideal ML" we refer to a ML with an ideal phase value at each lattice position and capable of converting 50% of the incident light from RCP to LCP before focusing LCP light. With ideal lens we refer to a lens with an ideal phase profile and with 100% transmission efficiency. All the conversion efficiency values are computed with the same integration radius $R_i = 17\,\mu m$.

**Discussion**

Similar to other adaptive optics approaches[52,54], the TR-ML proposed in this work is affected by chromatic aberrations. Our metalens is designed to operate at one wavelength of $\lambda = 0.632\ \mu m$ with an effective focal length $f \sim 150\ \mu m$ and an effective $NA = 0.1$. At this operating wavelength, focal length and NA, and assuming that the fractional change in the focal length is equal to the fractional change in the wavelength ($\Delta f/f = \Delta \lambda/\lambda$), the theoretical operation bandwidth[52,55] should be $\Delta \lambda \sim \pi/4 * \ln(2) * \lambda^2 / f\ NA^2 \sim 270\ nm$. However, a linear dependence of the phase on the wavelength might be not a good approximation because of the presence of resonance modes. Therefore, to ensure achromatic tunable metalens, different approaches recently demonstrated[8,56] should be specifically studied and combined for phase dispersion control over continuous bandwidth.

Our design principle enables also higher NA thermally tunable metalens. As the proof of concept, we performed the beam propagation resulting from the design of a thermally tunable metalens with $NA = 0.4$, radius $R = 4.55\ \mu m$ and a total focal length shift of 10% (from $10\ \mu m$ up to $11\ \mu m$). This design is based on the same nano-resonators set identified for our initial design and has been obtained following the same methodological approach. In general, the maximum focal length shift is limited by the tradeoff between the metalens NA and radius as well as the tradeoff between the metalens NA and the maximum thermal phase shift provided by the nano-resonators. Such trade-off also impacts the design of MLs with larger diameters, as they require a large maximum phase shift. However, selecting longer operation wavelengths, the ML design present a less stringent trade-off between maximum focal length shift and NA.

In this regard, although the thermally tunable metalens presented here has a small sub-millimeter aperture, our methods can be easily adapted to generate metalens designs with higher diameters while keeping the exact same set of nanofin resonators and parameter space. However, to reach high quality design while increasing the ML radius it is recommended either to decrease the total focal length variation requirement or to extend the parameter space by introducing other geometries to cover the range of the total phase thermal shift that will be required while increasing the ML radius. In particular, one possibility to increase the maximum phase shift is to combine both positive and negative phase shifts of the meta-atoms. Finally, it would be interesting to explore the use of a material with a higher thermo-optic coefficient, such as germanium[61,62], and to operate at infrared (telecom) wavelengths.

**Conclusions**

Overall, we have demonstrated a thermally-reconfigurable metalens based on thermo-optic effects in silicon nanofin resonators. Our design enables a continuous modulation of the focal length with close-to diffraction limited optical performance and competitive focusing efficiency. Our approach combines a geometrical phase approach with a temperature dependent phase shift to tune the lens properties using a spatially-uniform temperature input. Increasing the complexity of the meta-atom geometry and introducing more resonance modes[63] would offer a broader parameter space to search for TR-ML designs with advanced optical properties (high NA, high efficiency). Thus, we envision that coupling our approach with machine-learning algorithms[64,65] capable of identifying non-intuitive meta-atom shapes, which cover a large phase shift range with high conversion efficiencies, will significantly improve the performance of thermally-reconfigurable silicon metalenses. More broadly, the proposed design approach, based on geometric phase combined with thermo-optical effects, is expected to offer inspiration for the future realization of a broad class of active metasurfaces within the emerging field of thermo-nanophotonics.

## Methods

### Strehl ratio

Strehl ratio is defined as the ratio between the square of the electric field amplitude at the center of the designed intensity profile and the square of the electric field amplitude at the center of the ideal point spread function (PSF)[66–68]. It is given by the following equation:

$$S = \frac{I_{ML}(0,0)}{I_{th}(0,0)} = |\langle e^{i\phi}\rangle|^2 \approx e^{-\sigma^2} \tag{6}$$

where $\sigma = \sqrt{\langle(\phi - \bar{\phi})^2\rangle}$ is the root mean square of the deviation of the designed phase from the theoretical phase. A lens is typically considered to correspond to diffraction limited performance if the Strehl ratio exceeds $0.8$[66–68].

The root mean square deviation of the designed ML phase profile from the analytical phase profile results to be $\sigma[\phi(T_0)] = 2.5$ deg at $20°C$ and $\sigma[\phi(T_{\max})] = 4.3$ deg at $260°C$. These values result in a Strehl ratio of $S \sim 0.99$.

### Simulation of the silicon nano-resonators

The geometry of all nano-resonators was simulated with a fixed gap between the two nano-pillars ($g = 50\ nm$) and height ($h = 300\ nm$). The gap has been minimized to maximize the coupling and its value has been chosen based on fabrication considerations such as the resolution of e-beam lithography. The nano-pillar high was fixed at about $h \sim \lambda/2$ to ensure good mode coupling.

We simulated the interaction of circularly polarized light (electrical field $\boldsymbol{E}$) with our nano-scatter to extract both the phase $\phi$ and the transmission $T$ (coincident with the polarization conversion efficiency from RCP to LCP) of the electric field. We thus used the radio frequency (RF) module of COMSOL Multiphysics v5.5 to solve the Maxwell's equation:

$$\nabla \times (\nabla \times \boldsymbol{E}) - k_0^2 \epsilon_r \boldsymbol{E} = \boldsymbol{0} \tag{7}$$

where $\epsilon_r = (n - ik)^2 = n^2$, n and k are the real and the imaginary part of the refractive index $n$ respectively, $k_0 = 2\pi/\lambda$ is the wave number of free space and where only the constitutive relations for linear materials are considered.

Each silicon nanofins unit was simulated setting a periodic boundary condition along the transverse direction with respect to the propagation of light and a perfectly matched layer and input/output ports boundary conditions along the longitudinal direction.

We used the scattering parameter $S_{21}$, measured from the eigenmode expansion of the electromagnetic field at the output port 2, to extract both the transmission $T$ and phase $\phi$ of the electric field converted from right to left circularly polarized light:

$$\phi = -\arg(E^{out})$$
$$T = abs(E^{out})$$

where

$$S_{21} = \sqrt{\frac{Power\ deliverd\ to\ Port\ 2}{Power\ delivered\ to\ Port\ 1}}$$

and where the input RCP light incident on the input port 1 ($S_1$, S-parameter of incident wave) under sapphire substrate, and the output LCP component ($S_2$) at the exit port 2 are:

$$Port\ 1 \rightarrow E^{in,RCP} = \begin{pmatrix} 1 \\ i \end{pmatrix}$$
$$Port\ 2 \rightarrow E^{out,LCP} = \begin{pmatrix} 1 \\ -i \end{pmatrix}$$

Note that, while performing COMSOL simulation, we inverted both the sign of output phase and the input/output polarization definition compared to the ones used in the formalism described above for the phase profile design of our metalens. We are inverting the phase and polarization because in COMSOL the phase convention is based on:

$$\boldsymbol{E}^{CM}(x,y,z,\phi) = \boldsymbol{E_0}(x,y,z)e^{-i\phi} \qquad (8)$$

Thus, while in our formalism an increase of the phase value indicates retardation (delay accumulation) in phase profile, in COMSOL, a phase increase introduces an electrical field anticipation.

*Thermally tunable metalens design method*

The design of our metalens (ML) is based on three main steps written in Matlab.

In the first step, the software computes the maximum focal length variation and numerical aperture (NA) for a metalens characterized by the inputs set by the user. At this stage, the input data simulated in COMSOL are loaded and shown and the analytical phase profile based on the user input parameters is created and displayed. The metalens design optimization and the distribution of the different NF# geometries over the ML surface, are performed during the second step where the metalens layout is created and the Strehl ratio is computed. In the last step the Beam Propagation Method (BSM) is used to retrieve the actual ML focus profile and to extract the ML focus position, its FWHM and its depth of focus and its efficiency at each temperature. Lastly, the ML performance is compared with the behavior of an ideal. Further details can be found in the Supplementary note S1 and in the following paragraph.

*Beam Propagation Method (BPM)*

In this section, we present the derivation of the theory supporting the beam propagation method here implemented to study the behavior of a beam focused by our designed metalens[69,70]. All numerical calculations of the intensity profiles were performed in MATLAB environment.

When considering a plane wave incident on a homogeneous, isotropic, and linear medium, the electromagnetic wave equations reduce to the Helmholtz equation for $E$:

$$(\nabla^2 + k^2)\, E(x,y,z)e^{i\phi} = 0 \qquad (9)$$

In the paraxial limit, the two-dimensional Fourier transform of the electric field $E$ transverse to the propagation axis Z and at a certain fixed position z, (i.e. the $E$ field in the z = 0 plane) can be written as:

$$\widehat{E}(k_x, k_y; z) = \frac{1}{4\pi^2} \iint_{-\infty}^{\infty} E(x,y,z) e^{i\phi} e^{-i[k_x x + k_x x]}\, dx\, dy \qquad (10)$$

where $k_x, k_y$ are the spatial frequencies coordinates of the Cartesian transverse coordinates x and y and $\phi$ the phase value in the $(x, y, z)$ position.

Then, the inverse Fourier transform becomes:

$$E(x,y,z) = \iint_{-\infty}^{\infty} \widehat{E}(k_x, k_y; z) e^{i[k_x x + k_x x]}\, dk_x\, dk_y \qquad (11)$$

Knowing that:

$$k_z = \sqrt{(k^2 - k_x^2 - k_y^2)} \qquad (12)$$

and replacing the $E$ field expression in the Helmholtz equation with the Fourier representation just derived above, we find that the evolution of the field in the propagation direction Z, in the Fourier space corresponds to a multiplication for a factor $e^{\pm i k_z z}$:

$$\widehat{E}(k_x, k_y; z) = \widehat{E}(k_x, k_y; 0) e^{\pm i k_z z} \qquad (13)$$

In other words, the Fourier spectrum of $E$, in an arbitrary image plane at the z location, corresponds to the spectrum in the object plane ($z = 0$) multiply by the factor $e^{\pm i k_z z}$ where the sign '+' describes a wave propagating the forward positive direction $z > 0$ and '–' sign refers to a wave propagating in the negative half-space $z < 0$. Replacing the expression ( 14 ) into ( 12 ) we can express the electrical field as:

$$E(x,y,z) = \iint_{-\infty}^{\infty} \widehat{E}(k_x, k_y; 0) e^{i[k_x x + k_x x]} e^{\pm i k_z z} \, dk_x \, dk_y \quad (14)$$

where

$$\widehat{E}(k_x, k_y; 0) = \frac{1}{4\pi^2} \iint_{-\infty}^{\infty} E_0 e^{i\phi} e^{-i[k_x x + k_x x]} \, dx \, dy \quad (15)$$

is the Fourier transform of the electrical field at the ML plane ($z = 0$) and $\phi$ and $E_0$ and are the phase and the electrical field amplitude of each nano-resonator at the position $(x, y)$ on the $z = 0$ plane respectively. The intensity profile of the beam focused by the designed metalens can be finally retrieved from the squared of the electrical field propagation:

$$I(x, y, z) = |E(x, y, z)|^2 \quad (16)$$

*Geometrical phase*

The phase delay induced by dielectric nano-structures on the incident light can be controlled by tuning either the refractive index, the relative rotation or the resonances of the nano-structures. Both on- and off-resonance based approaches are strictly related to the operating wavelength, the refractive index of the utilized materials and the geometry and location of the nano-antennas. On the contrary, the relative rotation of the nano-pillars introduces a geometrical phase delay which depends only on the geometrical asymmetry of the nano-structures and on the polarization of the incident light.

According to the Pancharatnam-Berry geometric effect[48,49], circularly polarized light incident on a periodic layer done by subwavelength and anisotropic structures, with different orientations $\theta$ respect to the reference, is transmitted as the sum of two components: one with the same phase delay and the same handedness of the incident light and the other with a phase delay $\phi$ proportional to the rotation angle $\theta$ of the structure and with opposite handedness. In further detail, if the electric field incident on the nano-structure is circularly polarized, that is, $E^{in} = \begin{pmatrix} 1 \\ \pm i \end{pmatrix}$, then the output electric field is given by[48,71–74]:

$$E^{out} = \mathbf{M} \cdot E^{in} = \frac{t_l + t_s}{2} \begin{pmatrix} 1 \\ \pm i \end{pmatrix} + \frac{t_l - t_s}{2} e^{\pm i 2\theta} \begin{pmatrix} 1 \\ \mp i \end{pmatrix} \quad (17)$$

where the entire operation is represented by the matrix:

$$\mathbf{M} = \mathbf{R}(-\theta) \begin{pmatrix} t_l & 0 \\ 0 & t_s \end{pmatrix} \mathbf{R}(\theta) \quad (18)$$

where $t_l$ and $t_s$ are the complex transmittance coefficients corresponding to an incident light linearly polarized along the long and the short axis of the nanofin respectively, $\theta$ is the nano-structure rotation angle with respect to its long axis in the rotation, $\boldsymbol{R} = \begin{pmatrix} \cos\theta & \sin\theta \\ -\sin\theta & \cos\theta \end{pmatrix}$ is the rotation matrix and $\begin{pmatrix} 1 \\ \mp i \end{pmatrix}$ the eigenvectors associated to the eigenvalues $e^{\pm i\, 2\theta}$. The resulting phase delay of the light component with opposite handedness, thus, is $\phi = \pm 2\theta$.

The design of our ML relies on the electric field component converted from right-handed circularly polarized light $E^{in,RCP} = \begin{pmatrix} 1 \\ -i \end{pmatrix}$ to left-handed circularly polarized light $E^{out,LCP} = \frac{t_l - t_s}{2} e^{-i\, 2\theta} \begin{pmatrix} 1 \\ +i \end{pmatrix}$. The latter allows an average phase delay $\phi(\theta, T, \#) \sim -2\theta$ that can be exploited to cover the phase range needed for a thermally tunable focal length (**Figure 2c**).

**References**

1. Huang, Y.-W. *et al.* Aluminum Plasmonic Multicolor Meta-Hologram. *Nano Lett.* **15**, 3122–3127 (2015).

2. Sun, S. *et al.* High-Efficiency Broadband Anomalous Reflection by Gradient Meta-Surfaces. *Nano Lett.* **12**, 6223–6229 (2012).

3. Chen, W. T. *et al.* High-Efficiency Broadband Meta-Hologram with Polarization-Controlled Dual Images. *Nano Lett.* **14**, 225–230 (2014).

4. Kildishev, A. V., Boltasseva, A. & Shalaev, V. M. Planar Photonics with Metasurfaces. *Science* **339**, 1232009–1232009 (2013).

5. Arbabi, A., Horie, Y., Ball, A. J., Bagheri, M. & Faraon, A. Subwavelength-thick lenses with high numerical apertures and large efficiency based on high-contrast transmitarrays. *Nat. Commun.* **6**, 7069 (2015).

6. Khorasaninejad, M. *et al.* Metalenses at visible wavelengths: Diffraction-limited focusing and subwavelength resolution imaging. *Science* **352**, 1190–1194 (2016).

7. Wang, S. *et al.* A broadband achromatic metalens in the visible. *Nat. Nanotechnol.* **13**, 227–232 (2018).

8. Chen, W. T. *et al.* A broadband achromatic metalens for focusing and imaging in the visible. *Nat. Nanotechnol.* **13**, 220–226 (2018).

9. Chu, C. H. *et al.* Active dielectric metasurface based on phase-change medium. *Laser Photonics Rev.* **10**, 986–994 (2016).

10. Huang, Y.-W. *et al.* Gate-Tunable Conducting Oxide Metasurfaces. *Nano Lett.* **16**, 5319–5325 (2016).

11. Du, X. *et al.* Graphene-embedded broadband tunable metamaterial absorber in terahertz band. *J. Opt.* **22**, 015102 (2019).

12. Mou, N. *et al.* Large-scale, low-cost, broadband and tunable perfect optical absorber based on phase-change material. *Nanoscale* **12**, 5374–5379 (2020).

13. Leitis, A. *et al.* All-Dielectric Programmable Huygens' Metasurfaces. *Adv. Funct. Mater.* **30**, 1910259 (2020).

14. Mou, N. *et al.* Large-scale, low-cost, broadband and tunable perfect optical absorber based on phase-change material. *Nanoscale* **12**, 5374–5379 (2020).

15. Wang, Q. *et al.* Optically reconfigurable metasurfaces and photonic devices based on phase change materials. *Nat. Photonics* **10**, 60–65 (2016).

16. Kamali, S. M., Arbabi, E., Arbabi, A., Horie, Y. & Faraon, A. Highly tunable elastic dielectric metasurface lenses. *Laser Photonics Rev.* **10**, 1002–1008 (2016).

17. Ee, H.-S. & Agarwal, R. Tunable Metasurface and Flat Optical Zoom Lens on a Stretchable Substrate. *Nano Lett.* **16**, 2818–2823 (2016).

18. Malek, S. C., Ee, H.-S. & Agarwal, R. Strain Multiplexed Metasurface Holograms on a Stretchable Substrate. *Nano Lett.* **17**, 3641–3645 (2017).

19. She, A., Zhang, S., Shian, S., Clarke, D. R. & Capasso, F. Adaptive metalenses with simultaneous electrical control of focal length, astigmatism, and shift. *Sci. Adv.* **4**, eaap9957 (2018).

20. Rahmani, M. *et al.* Reversible Thermal Tuning of All-Dielectric Metasurfaces. *Adv. Funct. Mater.* **27**, 1700580 (2017).

21. Afridi, A. *et al.* Electrically Driven Varifocal Silicon Metalens. *ACS Photonics* 7 (2018).

22. Lewi, T., Butakov, N. A. & Schuller, J. A. Thermal tuning capabilities of semiconductor metasurface resonators. *Nanophotonics* **8**, 331–338 (2019).

23. Iyer, P. P., DeCrescent, R. A., Lewi, T., Antonellis, N. & Schuller, J. A. Uniform Thermo-Optic Tunability of Dielectric Metalenses. *Phys. Rev. Appl.* **10**, 044029 (2018).

24. Brar, V. W. *et al.* Electronic modulation of infrared radiation in graphene plasmonic resonators. *Nat. Commun.* **6**, 7032 (2015).

25. Ding, P. *et al.* Graphene aperture-based metalens for dynamic focusing of terahertz waves. *Opt. Express* **26**, 28038 (2018).

26. Du, X. *et al.* Graphene-embedded broadband tunable metamaterial absorber in terahertz band. *J. Opt.* **22**, 015102 (2019).

27. Chen, W. T. *et al.* Immersion Meta-Lenses at Visible Wavelengths for Nanoscale Imaging. *Nano Lett.* **17**, 3188–3194 (2017).

28. Guo, Q. *et al.* Compact single-shot metalens depth sensors inspired by eyes of jumping spiders. *Proc. Natl. Acad. Sci.* **116**, 22959–22965 (2019).

29. Lin, R. J. *et al.* Achromatic metalens array for full-colour light-field imaging. *Nat. Nanotechnol.* **14**, 227–231 (2019).

30. Lee, G.-Y. *et al.* Metasurface eyepiece for augmented reality. *Nat. Commun.* **9**, 4562 (2018).

31. Lin, R. J. *et al.* Achromatic metalens array for full-colour light-field imaging. *Nat. Nanotechnol.* **14**, 227–231 (2019).

32. Engelberg, J. & Levy, U. The advantages of metalenses over diffractive lenses. *Nat. Commun.* **11**, 1991 (2020).

33. Wei, S. *et al.* A Varifocal Graphene Metalens for Broadband Zoom Imaging Covering the Entire Visible Region. *ACS Nano* (2021) doi:10.1021/acsnano.0c09395.

34. She, A., Zhang, S., Shian, S., Clarke, D. R. & Capasso, F. Adaptive metalenses with simultaneous electrical control of focal length, astigmatism, and shift. *Sci. Adv.* **4**, eaap9957 (2018).

35. Zograf, G. P., Petrov, M. I., Makarov, S. V. & Kivshar, Y. S. All-dielectric thermonanophotonics. *ArXiv210401964 Phys.* (2021).

36. van de Groep, J. *et al.* Exciton resonance tuning of an atomically thin lens. *Nat. Photonics* **14**, 426–430 (2020).

37. Dias, E. J. C., Yu, R. & García de Abajo, F. J. Thermal manipulation of plasmons in atomically thin films. *Light Sci. Appl.* **9**, 87 (2020).

38. Park, S. *et al.* Electrically focus-tuneable ultrathin lens for high-resolution square subpixels. *Light Sci. Appl.* **9**, 98 (2020).

39. Tsoulos, T. V. & Tagliabue, G. Self-induced thermo-optical effects in silicon and germanium dielectric nanoresonators. *Nanophotonics* **9**, 3849–3861 (2020).

40. Chen, W. T., Zhu, A. Y. & Capasso, F. Flat optics with dispersion-engineered metasurfaces. *Nat. Rev. Mater.* **5**, 604–620 (2020).

41. Park, S. *et al.* Electrically focus-tuneable ultrathin lens for high-resolution square subpixels. *Light Sci. Appl.* **9**, 98 (2020).

42. Afridi, A. *et al.* Electrically Driven Varifocal Silicon Metalens. *ACS Photonics* 7 (2018).

43. Liu, M., Fan, Q., Yu, L. & Xu, T. Polarization-independent infrared micro-lens array based on all-silicon metasurfaces. *Opt. Express* **27**, 10738 (2019).

44. Lu, X. *et al.* Broadband achromatic metasurfaces for sub-diffraction focusing in the visible. *Opt. Express* **29**, 5947–5958 (2021).

45. Rahmani, M. *et al.* Reversible Thermal Tuning of All-Dielectric Metasurfaces. *Adv. Funct. Mater.* **27**, 1700580 (2017).

46. Iyer, P. P., DeCrescent, R. A., Lewi, T., Antonellis, N. & Schuller, J. A. Uniform Thermo-Optic Tunability of Dielectric Metalenses. *Phys. Rev. Appl.* **10**, 044029 (2018).

47. Chen, M. K. *et al.* Principles, Functions, and Applications of Optical Meta-Lens. *Adv. Opt. Mater.* **9**, 2001414 (2021).

48. Gori, F. Measuring Stokes parameters by means of a polarization grating. *Opt. Lett.* **24**, 584 (1999).

49. Roux, F. S. Geometric phase lens. *J. Opt. Soc. Am. A* **23**, 476 (2006).

50. Chen, W. T., Zhu, A. Y., Sisler, J., Bharwani, Z. & Capasso, F. A broadband achromatic polarization-insensitive metalens consisting of anisotropic nanostructures. *Nat. Commun.* **10**, 355 (2019).

51. Ruffieux, P., Scharf, T., Herzig, H. P., Völkel, R. & Weible, K. J. On the chromatic aberration of microlenses. *Opt. Express* **14**, 4687 (2006).

52. Arbabi, E. *et al.* MEMS-tunable dielectric metasurface lens. *Nat. Commun.* **9**, 812 (2018).

53. Bosch, M. *et al.* Electrically Actuated Varifocal Lens Based on Liquid-Crystal-Embedded Dielectric Metasurfaces. *Nano Lett.* (2021) doi:10.1021/acs.nanolett.1c00356.

54. Klopfer, E., Lawrence, M., Barton, D. R., Dixon, J. & Dionne, J. A. Dynamic Focusing with High-Quality-Factor Metalenses. *Nano Lett.* **20**, 5127–5132 (2020).

55. Arbabi, A. *et al.* Miniature optical planar camera based on a wide-angle metasurface doublet corrected for monochromatic aberrations. *Nat. Commun.* **7**, 13682 (2016).

56. Zhu, A. Y. *et al.* Compact Aberration-Corrected Spectrometers in the Visible Using Dispersion-Tailored Metasurfaces. *Adv. Opt. Mater.* **7**, 1801144 (2019).

57. Yang, Y., Kravchenko, I. I., Briggs, D. P. & Valentine, J. High Quality Factor Fano-Resonant All-Dielectric Metamaterials. *ArXiv14053901 Phys.* (2014).

58. Yang, Z.-J., Zhao, Q. & He, J. Fano interferences of electromagnetic modes in dielectric nanoblock dimers. *J. Appl. Phys.* **125**, 063103 (2019).

59. Su, Z., Song, K., Yin, J. & Zhao, X. Metasurface with interfering Fano resonance: manipulating transmission wave with high efficiency. *Opt. Lett.* **42**, 2366 (2017).

60. Yan, C., Yang, K.-Y. & Martin, O. J. F. Fano-resonance-assisted metasurface for color routing. *Light Sci. Appl.* **6**, e17017–e17017 (2017).

61. Bosch, M., Shcherbakov, M. R., Fan, Z. & Shvets, G. Polarization states synthesizer based on a thermo-optic dielectric metasurface. *J. Appl. Phys.* **126**, 073102 (2019).

62. Viña, L., Logothetidis, S. & Cardona, M. Temperature dependence of the dielectric function of germanium. *Phys. Rev. B* **30**, 1979–1991 (1984).

63. Hsiao, H.-H. *et al.* Integrated Resonant Unit of Metasurfaces for Broadband Efficiency and Phase Manipulation. *Adv. Opt. Mater.* **6**, 1800031 (2018).

64. An, S. *et al.* Deep learning modeling approach for metasurfaces with high degrees of freedom. *Opt. Express* **28**, 31932 (2020).

65. Zhelyeznyakov, M. V., Brunton, S. L. & Majumdar, A. Deep learning to accelerate Maxwell's equations for inverse design of dielectric metasurfaces. *ArXiv200810632 Phys.* (2020).

66. Ottevaere, H. & Thienpont, H. OPTICAL MICROLENSES. in *Encyclopedia of Modern Optics* (ed. Guenther, R. D.) 21–43 (Elsevier, 2005). doi:10.1016/B0-12-369395-0/00923-4.

67. K. L. Chuang, K. K. *Statistical Analyis of the 70 Meter Antenna Surface Distorsion*. 29–35 https://tmo.jpl.nasa.gov/progress_report/42-88/88E.PDF (1986).

68. Mahajan, V. N. Strehl ratio for primary aberrations in terms of their aberration variance. *JOSA* **73**, 860–861 (1983).

69. Kamilov, U. S. *et al.* Optical Tomographic Image Reconstruction Based on Beam Propagation and Sparse Regularization. *IEEE Trans. Comput. Imaging* **2**, 59–70 (2016).

70. Feit, M. D. & Fleck, J. A. Beam nonparaxiality, filament formation, and beam breakup in the self-focusing of optical beams. *JOSA B* **5**, 633–640 (1988).

71. S. Pancharatnam. Generalized theory of interference, and its applications. *Proc Indian Acad Sci Sect A* **44**, 247–262 (1956).

72. Bomzon, Z., Biener, G., Kleiner, V. & Hasman, E. Space-variant Pancharatnam–Berry phase optical elements with computer-generated subwavelength gratings. *Opt. Lett.* **27**, 1141–1143 (2002).

73. Roux, F. S. Geometric phase lens. *J. Opt. Soc. Am. A* **23**, 476 (2006).

74. Zhan, T., Xiong, J., Lee, Y.-H. & Wu, S.-T. Polarization-independent Pancharatnam-Berry phase lens system. *Opt. Express* **26**, 35026–35033 (2018).

75. Lalanne, P. Waveguiding in blazed-binary diffractive elements. *J. Opt. Soc. Am. A* **16**, 2517 (1999).

76. Vuye, G. *et al.* Temperature dependence of the dielectric function of silicon using in situ spectroscopic ellipsometry. *Thin Solid Films* **233**, 166–170 (1993).

77. Rybin, M. V. *et al.* High- Q Supercavity Modes in Subwavelength Dielectric Resonators. *Phys. Rev. Lett.* **119**, 243901 (2017).

78. Yang, Y. *et al.* Multimode directionality in all-dielectric metasurfaces. *Phys. Rev. B* **95**, 165426 (2017).